\documentclass[prb,showpacs,amsmath,amssymb,amstext,amsfont]{revtex4}
\usepackage{graphicx}
\usepackage{bm}

\def\br{\bm{r}}
\def\bk{\bm{k}}

\def\bmeta{\bm{\eta}}
\def\im{\,\mathrm{Im}}

\def\sign{\,\mathrm{sign}\,}
\def\tr{\,\mathrm{tr}}
\def\Tr{\,\mathrm{Tr}}

\begin{document}

\title{Domain walls in chiral $p$-wave superconductors: Quasiparticle spectrum and dynamics}

\author{K. V. Samokhin}

\affiliation{Department of Physics, Brock University, St.Catharines, Ontario, Canada L2S 3A1}
\date{\today}

\begin{abstract}
We calculate microscopically the viscous friction coefficient and the effective mass of domain walls separating regions of opposite chirality in $p$-wave superconductors with $k_x\pm ik_y$ order parameter. The domain wall viscosity and inertia
are determined by the transitions between different Bogoliubov quasiparticle states induced by the domain wall motion. As a by-product, we present a detailed analysis of the quasiparticle spectrum, both bound and scattering, 
in the presence of a general domain wall with an arbitrary phase difference between the domains.
\end{abstract}


\maketitle

\section{Introduction}
\label{sec: Intro}

The properties of topological superconductors and superfluids have attracted a considerable interest recently. The defining feature of these systems is that, while the fermionic excitations in the bulk are fully gapped,
nontrivial topology of the order parameter manifests itself in the existence of gapless quasiparticles localized near the order parameter inhomogeneities, such as Abrikosov vortices, domain walls, or the sample boundaries.
One of the most studied examples is the chiral $p$-wave spin-triplet state, whose experimental realizations include the superconducting state of Sr$_2$RuO$_4$ (Ref. \onlinecite{SrRuO}) or
thin films of superfluid ${}^3$He-$A$ (Ref. \onlinecite{Vol92}). The chiral $p$-wave state in fermionic superfluids is closely related to the Moore-Read Pfaffian quantum Hall state.\cite{RG00}

The momentum-space order parameter of a chiral $p$-wave superconductor is proportional to $k_x\pm ik_y$. 
In the absence of external magnetic field, it is two-fold degenerate: the states $k_x+ik_y$ and $k_x-ik_y$, which are obtained from each other by time reversal, have the same energy. Therefore, superconducting states with opposite chiralities
separated by domain walls (DW) might form in different parts of the system. Indeed, there is evidence of the DW existence in Sr$_2$RuO$_4$ (Refs. \onlinecite{Kidwin06} and \onlinecite{Kambara08}) and also in slabs of superfluid ${}^3$He 
(Ref. \onlinecite{WWG04}). The DW formation costs gradient energy due to the spatial variation of
the order parameter. In contrast to ferromagnets, which break up into domains in order to minimize the net magnetic moment, there is no similarly compelling energy reason
in a neutral superfluid. One possible mechanism is that domains are spontaneously formed upon cooling across the phase transition due to the sample inhomogeneity. Alternatively, an increase in the gradient energy
might be compensated by the creation of low-energy quasiparticles bound to the DW, which is particularly effective in one-dimensional systems.\cite{KY02}

While the static properties of the DWs in various realizations of the chiral $p$-wave state have been extensively studied,\cite{Ho84,VG85,Naka86,BK88,SRU89,MS99,SA99,SR04,Liu05,LS09,BS10} their dynamics has received comparatively little theoretical attention,
see Ref. \onlinecite{Sam11}. The motion of a different type of planar defects, namely, an interface between the $A$ and $B$ phases of superfluid ${}^3$He was studied in Refs. \onlinecite{AB-boundary-1} and \onlinecite{AB-boundary-2}, where 
it was pointed out that the scattering of Bogoliubov quasiparticles by the moving interface results in an effective friction force. Similar ideas can be applied in our case as well. The DW, which is assumed to be moving uniformly as a whole, suffers viscous 
friction and acquires mass due to its interaction with fermionic quasiparticles, see Sec. \ref{sec: Model}. To obtain the DW dynamic characteristics we employ the effective bosonic action formalism, see Sec. \ref{sec: S-eff derivation}. 
The Gaussian effective action for the DW essentially depends on the Bogoliubov quasiparticle spectrum, both bound and scattering states, 
in the presence of a static DW. The latter is studied in detail for a general DW structure in Sec. \ref{sec: QP spectrum}. Although some bits and pieces about the properties of the DW quasiparticle spectrum can be found scattered in the literature, 
we believe it is useful to present a complete picture in one place. Some of the more technical details are discussed in four appendices. Finally, the DW friction coefficient and the zero-temperature effective mass are calculated in Sec. \ref{sec: M and eta}.
Throughout the paper we use the units in which $\hbar=k_B=c=1$.

\section{The model}
\label{sec: Model}

Let us consider a two-dimensional triplet $p$-wave fermionic superfluid or superconductor. The gap function is a spin matrix given by $i(\hat{\bm{\sigma}}\bm{d})\hat\sigma_2$, where
$\bm{d}=\hat z(\eta_1k_x+\eta_2k_y)/k_F$ describes triplet pairing, $k_F$ is the Fermi wavevector, and $\hat{\bm{\sigma}}$ are the Pauli matrices.\cite{Book} The order parameter $\bmeta=(\eta_1,\eta_2)$ is characterized by two planar components, 
which can depend on coordinates and, in the dynamic case, time. We assume that the band dispersion is isotropic: $\xi(\bk)=(\bk^2-k_F^2)/2m^*$,
with the effective mass $m^*$ (our results can be straightforwardly generalized for the case of anisotropic dispersion). We also neglect disorder, as well as the effects related to the electric charges, such as the Meissner screening of 
the external or internal magnetic fields. 

For a static DW, one can choose the $x$-axis along the normal and write the order parameter as 
\begin{equation}
\label{DW-general-structure}
  \bmeta=(|\eta_1|,|\eta_2|e^{-i\gamma})e^{i\phi},
\end{equation}
where the amplitudes of the components, the relative phase $\gamma$, and the common phase $\phi$ all depend on $x$. 
Assuming that the most stable superconducting state in the bulk is described by one of the two degenerate chiral states $\bmeta\propto(1,\pm i)$ and allowing for a nonzero phase difference between the two domains, 
we have the following expression for the order parameter asymptotics far from the DW:
\begin{equation}
\label{DW-model}
  \begin{array}{ll}
  \bmeta(x)=\Delta_0(1,i),\qquad & x\to-\infty\\
  \bmeta(x)=\Delta_0e^{i\chi}(1,-i),\qquad & x\to+\infty.
  \end{array}
\end{equation}
Here $\chi$ is a parameter which depends on the microscopic details ($0\leq\chi\leq\pi$). Its value is fixed by the condition of vanishing supercurrent across the DW, see a discussion of this point in Appendix \ref{sec: GL description}. 

An exact analytical expression for the DW structure is not available and a variety of approximations have been proposed in the literature. For instance, the amplitudes of both components can be put
constant: $|\eta_1|=|\eta_2|=\Delta_0$ (Ref. \onlinecite{VG85}). Alternatively, one can assume constant phases: $\bmeta=\Delta_0(1,if)$ or $\bmeta=\Delta_0(if,1)$, where a real function $f(x)$ varies between $1$ at $x=-\infty$ and $-1$ at $x=\infty$ 
(Refs. \onlinecite{Vol92,Ho84,Naka86}, and \onlinecite{BK88}). Other possibilities include $\bmeta=\Delta_0(\cos\Theta,i\sin\Theta)$, where $\Theta(x)$ varies between $0$ and $\pi$ (Refs. \onlinecite{Liu05} and \onlinecite{LS09}), and 
$(\eta_+,\eta_-)=\Delta_0(e^{i\phi_+}\cos\zeta,e^{i\phi_-}\sin\zeta)$, where $\eta_\pm=(\eta_1\mp i\eta_2)/\sqrt{2}$, $\phi_\pm$ are parameters, and $\zeta(x)$ varies
between $0$ and $\pi/2$ (Refs. \onlinecite{SA99} and \onlinecite{BS10}). In all cases, the DW order parameter variation occurs within a region of width $\xi_d$ around the origin $x=0$, with $\xi_d$ being the DW thickness. 
The precise way in which the DW order parameter varies between the asymptotics given by Eq. (\ref{DW-model}) is not important for our purposes. 

We are interested in the motion of the DW as a whole. Such a motion can be caused, for instance, by the (extremely weak) interaction of an external magnetic field with the orbital moment of the Cooper pairs.\cite{Legg77}
The direction of the latter is given by the unit vector $\bm{l}=i(\bmeta^*\times\bmeta)/|\bmeta^*\times\bmeta|$, which takes opposite values in the two domains [according to Eq. (\ref{DW-general-structure}), $\bm{l}=\hat z\sign(\sin\gamma)$], 
thus creating a transverse force on the DW. For a small driving force one can expect a linear relation between the DW velocity and the force. According to the fundamental principles of nonequilibrium statistical mechanics, one
can express linear-response kinetic coefficients, in particular, the DW viscous friction, in terms of the equilibrium fluctuation properties. It is legitimate, 
therefore, to use the Matsubara formalism with the magnetic field and the driving force both set to zero. In the absence of external field, the controversy about the magnitude of the orbital moment and the lack of its local definition, see 
Refs. \onlinecite{Book,Kita98}, and \onlinecite{BS06}, is not relevant for our problem.

In the bosonic effective action formalism,\cite{Pop91} the order parameter becomes a two-component dynamical field $\bm{\eta}(\br,\tau)$, which depends not only on coordinates, but also on the Matsubara time $\tau$. We use the following ansatz for a moving DW:
\begin{equation}
\label{moving DW}
    \bmeta(\br,\tau)=\bmeta[x-X(\tau)],
\end{equation}
where $\bmeta(x)$ is the mean-field order parameter of a static DW and $X(\tau)$ is the macroscopic coordinate describing a uniform displacement of the DW. The effective action can be expressed in terms of $X$ as follows: 
\begin{equation}
\label{S-eff-2}
    S_{eff}[X]=S_0+T\sum_m{\cal K}(\nu_m)X(\nu_m)X(-\nu_m),
\end{equation}
where $S_0$ is the mean-field action for the static DW, the second term is the DW dynamic action in the Gaussian approximation, and $\nu_m=2m\pi T$ is the bosonic Matsubara frequency. Due to the translational invariance of the system, a static, 
i.e. time-independent, displacement would not affect the action, therefore, ${\cal K}(\nu_m=0)=0$.  

One can view the DW as a macroscopic quantum object interacting with the equilibrium bath of fermionic quasiparticles. Its motion induces transitions between the quasiparticle states, which result in the effective DW friction and inertia.  
If the leading term in the frequency expansion of the kernel ${\cal K}(\nu_m)$ is given by $\eta|\nu_m|$, then $\eta$ can be interpreted, in the spirit of the Caldeira-Leggett model,\cite{CL81} as the viscous friction coefficient, 
while a term quadratic in $\nu_m$ yields the inertial mass of the DW. The frequency is assumed to be small compared to the gap amplitude, so that the DW order parameter profile is not deformed during its motion. 
Note that there are different ways to define the effective mass of a topological defect in superconductor and superfluids, discussed mostly in the context of the Abrikosov vortex dynamics, see Ref. \onlinecite{vortex-mass}. 
Our approach, based on the Matsubara effective action, is similar to the one developed for the vortex dynamics in Ref. \onlinecite{vortex-action}. In general, the dynamics of any stable inhomogeneous order parameter structure can be analyzed 
in this way.\cite{WWS96}

\section{Derivation of the effective action}
\label{sec: S-eff derivation}

Our investigation of the DW dynamics is based on the bosonic effective action for a chiral $p$-wave superconductor. The standard procedure, which involves integrating out the fermionic degrees of freedom,\cite{Pop91} yields the following 
expression for the action: 
\begin{equation}
\label{S-eff}
    S_{eff}=\frac{1}{V}\int_0^\beta d\tau\int d\br|\bmeta(\br,\tau)|^2-\frac{1}{2}\Tr\ln{\cal G}^{-1},
\end{equation}
where $V>0$ is the coupling constant of the triplet $p$-wave pairing channel, $\beta$ is the inverse temperature, and
\begin{equation}
\label{G-1}
    {\cal G}^{-1}=\left(\begin{array}{cc}
        -\partial_\tau-\hat\xi & -\hat\Delta(\br,\tau) \\
        -\hat\Delta^\dagger(\br,\tau) & -\partial_\tau+\hat\xi \\
    \end{array}\right).
\end{equation}
Here $\hat\xi=\xi(\hat{\bk})$, $\hat\Delta(\br,\tau)=\hat\sigma_1[\eta_1(\br,\tau)\hat k_x+\eta_2(\br,\tau)\hat k_y]/k_F$, and $\hat{\bk}=-i\bm{\nabla}$.
For a DW moving as a whole, see Eq. (\ref{moving DW}), we have $\hat\Delta(\br,\tau)=\hat\Delta_0[x-X(\tau)]$, where
\begin{equation}
\label{Delta_0}
    \hat\Delta_0(x)=\hat\sigma_1\frac{\eta_1(x)\hat k_x+\eta_2(x)\hat k_y}{k_F}
\end{equation}
corresponds to the static DW, with the order parameter components given by Eq. (\ref{DW-general-structure}). 

The first term in the effective action (\ref{S-eff}) does not depend on $X$, while the second one can be expanded in powers of the displacement, using ${\cal G}^{-1}={\cal G}_0^{-1}-\Sigma$, where
\begin{equation}
\label{G0}
    {\cal G}_0(\br_1,\br_2;\omega_n)=\sum_j\frac{\langle\br_1|j\rangle\langle j|\br_2\rangle}{i\omega_n-E_j}
\end{equation}
is the Green's function at $X=0$. Here $\omega_n=(2n+1)\pi T$ is the fermionic Matsubara frequency, $|j\rangle$ and $E_j$ are the eigenfunctions and eigenvalues of the $4\times 4$ Bogoliubov-de Gennes (BdG) Hamiltonian for the static DW, which has
the following form:
\begin{equation}
\label{H_0}
	{\cal H}_0=\left(\begin{array}{cc}
		\hat\xi & \hat\Delta_0\\
		\hat\Delta^\dagger_0 & -\hat\xi
	\end{array}\right).
\end{equation}
The self-energy function $\Sigma$ describes the effects of the DW displacement. At the Gaussian level, we keep only the terms of the first and second order in $X$ in its expansion: $\Sigma=\Sigma_1+\Sigma_2$, where $\Sigma_1=-X{\cal H}_1$, with 
\begin{equation}
\label{cal H_1}
  {\cal H}_1=\left(\begin{array}{cc}
              0 & \nabla_x\hat\Delta_0 \\      
              \nabla_x\hat\Delta_0^\dagger & 0
	      \end{array}\right)
\end{equation}
and $\Sigma_2=(X^2/2){\cal H}_2$, with
$$
  {\cal H}_2=\left(\begin{array}{cc}
              0 & \nabla^2_x\hat\Delta_0 \\      
              \nabla^2_x\hat\Delta_0^\dagger & 0
	      \end{array}\right).
$$

Subsequent calculations are facilitated by two identities:
\begin{eqnarray}
\label{W-identity-1}
  &&\langle i|{\cal H}_1|j\rangle=-(E_i-E_j)W_{ij},\\
\label{W-identity-2}
  &&\langle i|{\cal H}_2|j\rangle=\sum_k(E_i+E_j-2E_k)W_{ik}W_{kj},
\end{eqnarray}
where $W_{ij}=\langle i|\nabla_x|j\rangle$ are the matrix elements of the generator of the DW translations. 
The first identity follows immediately from the expressions ${\cal H}_1=[\nabla_x,{\cal H}_0]$ and $\langle i|[\nabla_x,{\cal H}_0]|j\rangle=(E_j-E_i)\langle i|\nabla_x|j\rangle$.
To prove the second identity, we observe that ${\cal H}_2=[\nabla_x,[\nabla_x,{\cal H}_0]]$. On the other hand,
\begin{eqnarray*}
  &&\langle i|[\nabla_x,[\nabla_x,{\cal H}_0]]|j\rangle\\
  &&\quad=(E_i+E_j)\langle i|\nabla_x^2|j\rangle-2\langle i|\nabla_x{\cal H}_0\nabla_x|j\rangle\\
  &&\quad=(E_i+E_j)\sum_kW_{ik}W_{kj}-2\sum_k E_kW_{ik}W_{kj}.
\end{eqnarray*}
The last line here is obtained using the completeness of the eigenfunctions.

Inserting Eqs. (\ref{G0}), (\ref{W-identity-1}), and (\ref{W-identity-2}) into the expansion of $\Tr\ln{\cal G}^{-1}$, it is straightforward to show that the terms linear in $X$ vanish, while the quadratic terms can be collected into 
the expression (\ref{S-eff-2}) for the Gaussian action. The kernel is given by ${\cal K}={\cal K}_1+{\cal K}_2$, where 
\begin{eqnarray*}
  &&{\cal K}_1=\frac{1}{4}\Tr(\Sigma_1{\cal G}_0\Sigma_1{\cal G}_0)=\frac{1}{4}T\sum_n\sum_{ij}\frac{(E_i-E_j)^2}{(i\omega_n+i\nu_m-E_i)(i\omega_n-E_j)}|W_{ij}|^2,\\
  &&{\cal K}_2=\frac{1}{2}\Tr(\Sigma_2{\cal G}_0)=-\frac{1}{2}T\sum_n\sum_{ij}\frac{E_i-E_j}{i\omega_n-E_i}|W_{ij}|^2.
\end{eqnarray*}
Since ${\cal K}_2=-{\cal K}_1(\nu_m=0)$, the kernel vanishes for a stationary DW, as expected. Calculating the fermionic Matsubara sums, we arrive at the following result:
\begin{eqnarray}
\label{K-gen}
  {\cal K}(\nu_m)=\frac{1}{4}\sum_{ij}\frac{i\nu_m(E_i-E_j)}{E_i-E_j-i\nu_m}[f(E_i)-f(E_j)]|W_{ij}|^2\nonumber\\
  =\frac{\nu_m^2}{4}\sum_{ij}\frac{(E_i-E_j)[f(E_j)-f(E_i)]}{(E_i-E_j)^2+\nu_m^2}|W_{ij}|^2
\end{eqnarray}
where $f(E)=(e^{\beta E}+1)^{-1}$ is the Fermi function. It is easy to see that the transitions between the BdG eigenstates corresponding to the same energy do not contribute to ${\cal K}(\nu_m)$. 

We note that one could also arrive at  Eqs. (\ref{S-eff-2}) and (\ref{K-gen}) via a somewhat shorter route, using a change of coordinates
$x-X(\tau)=x'$, $y=y'$, $\tau=\tau'$, to transform into the reference frame co-moving with the DW. In this way, the invariance of the action under a static displacement of the DW is manifest from the beginning. 
A drawback of this approach is that the abovementioned transformation implies periodic boundary conditions for the fermionic wave functions as well as for the order parameter, which are actually inconsistent with a single DW.

It is convenient to represent the Gaussian kernel (\ref{K-gen}) in the following form:
\begin{equation}
\label{K-gen-int}
    {\cal K}(\nu_m)=\frac{\nu_m^2}{4}\int d\epsilon\,d\epsilon'\frac{(\epsilon-\epsilon')[f(\epsilon')-f(\epsilon)]}{(\epsilon-\epsilon')^2+\nu_m^2}{\cal N}(\epsilon,\epsilon'),
\end{equation}
where ${\cal N}(\epsilon,\epsilon')=\sum_{ij}|W_{ij}|^2\delta(\epsilon-E_i)\delta(\epsilon'-E_j)$, satisfying ${\cal N}(\epsilon,\epsilon')={\cal N}(\epsilon',\epsilon)$.
To avoid dealing with ill-defined expressions for the matrix elements involving the bulk quasiparticle states, see Sec. \ref{sec: QP spectrum} below, one can use the idenity (\ref{W-identity-1}) to obtain
\begin{equation}
\label{cal N def}
    {\cal N}(\epsilon,\epsilon')=\frac{1}{(\epsilon-\epsilon')^2}\sum_{ij}|\langle i|{\cal H}_1|j\rangle|^2\delta(\epsilon-E_i)\delta(\epsilon'-E_j),
\end{equation}
where the summation is performed over the pairs of the eigenstates of ${\cal H}_0$, satisfying $E_i\neq E_j$. 
The matrix elements of ${\cal H}_1$ are well defined, because the order parameter derivatives are nonzero only in the vicinity of the DW.

\section{Quasiparticle spectrum}
\label{sec: QP spectrum}

In order to calculate ${\cal N}(\epsilon,\epsilon')$, we need to know the quasiparticle spectrum for the static DW. In the absence of external magnetic field, the $4\times 4$ BdG Hamiltonian (\ref{H_0})
can be written as a direct sum of two identical $2\times 2$ Hamiltonians, labelled by the spin projection $\sigma=\uparrow,\downarrow$. From this point on we drop the spin index, restoring it only in the final expressions. 
For a DW parallel to the $y$-axis, the two-component wave function for each spin projection can be written as $e^{ik_yy}\Psi(x)$, where $\Psi(x)$ satisfies the equation
\begin{equation}
\label{BdG-eq-MF}
	\left(\begin{array}{cc}
		\dfrac{\hat k_x^2-k_0^2}{2m^*} & \Delta_0(x)\\
		\Delta^\dagger_0(x) & -\dfrac{\hat k_x^2-k_0^2}{2m^*}
	\end{array}\right)
	\Psi=E\Psi.
\end{equation}
Here $k_0=\sqrt{k_F^2-k_y^2}$ and $\Delta_0(x)=\eta_1(x)(\hat k_x/k_F)+\eta_2(x)(k_y/k_F)$.

Since the superconducting order parameter varies slowly on the scale of the inverse Fermi wavevector, one can use the semiclassical, or Andreev, approximation\cite{And64} and seek the quasiparticle wave functions in the form
$\Psi(x)=e^{ik_xx}\psi(x)$, where $k_x=\pm k_0$. The slowly-varying envelope function $\psi=(u,v)^T$ has the electron-like ($u$) and hole-like ($v$) components, which are found by solving the Andreev equation
\begin{equation}
\label{And-eq}
	\left(\begin{array}{cc}
		-iv_{F,x}\nabla_x & \Delta_{\bk_F}(x)\\
		\Delta^*_{\bk_F}(x) & iv_{F,x}\nabla_x
	\end{array}\right)\psi=E\psi.
\end{equation}
The Fermi wavevector $\bk_F\equiv (k_x,k_y)=k_F(\cos\theta,\sin\theta)$ defines the direction of semiclassical propagation of quasiparticles, along which the DW order parameter is given by
\begin{equation}
\label{Andreev-OP}
  \Delta_{\bk_F}(x)=\eta_1(x)\cos\theta+\eta_2(x)\sin\theta,
\end{equation}
and $v_{F,x}=v_F\cos\theta$ ($v_F=k_F/m^*$). Different models for the DW structure, see Sec. \ref{sec: Model}, result in different semiclassical order parameters. 
However, the asymptotic values of $\Delta_{\bk_F}(x)$ are fixed as follows:
\begin{equation}
\label{Delta-pm}
  \begin{array}{l}
    \Delta_-\equiv\Delta_{\bk_F}(x\ll-\xi_d)=\Delta_0e^{i\theta},\\
    \Delta_+\equiv\Delta_{\bk_F}(x\gg\xi_d)=\Delta_0e^{i\chi}e^{-i\theta},
  \end{array}
\end{equation}
according to Eq. (\ref{DW-model}).

Since $\Delta_{-\bk_F}(x)=-\Delta_{\bk_F}(x)$, the Andreev Hamiltonian defined by Eq. (\ref{And-eq}) has the property $H_{-\bk_F}=-H_{\bk_F}$. Therefore, the quasiparticle spectrum is electron-hole symmetric, in the following sense:
if $E$ is an eigenvalue of $H_{\bk_F}$, with the eigenfunction given by $\psi$, then $-E$ is an eigenvalue of $H_{-\bk_F}$, with the same eigenfunction.

At given $\bk_F$, the spectrum of Eq. (\ref{And-eq}) consists of a continuum of scattering states in the bulk, with $|E|\geq\Delta_0$, and also discrete bound states (called the Andreev bound states, or ABS's) with the energies $|E|<\Delta_0$. 
Below we discuss general properties of the spectrum which are independent of a particular choice for the DW structure.

\subsection{Scattering states}
\label{sec: scattering states}

At each energy satisfying $|E|\geq\Delta_0$, there are two scattering states, labelled by $L$ and $R$, corresponding to the two possible directions of propagation of the incident Andreev modes. Their wave functions
can be found explicitly only far from the DW, i.e. at $|x|\gg\xi_d$, where the order parameter is uniform. We have
\begin{eqnarray}
\label{EL}
    \psi_{E,L}(x)=C\left\{
    \begin{array}{l}
      \alpha_q^{(-)}e^{iqx}+r_L\alpha_{-q}^{(-)}e^{-iqx},\qquad  x\ll-\xi_d \medskip \\
      t_L\alpha_q^{(+)}e^{iqx},\qquad x\gg\xi_d
    \end{array}\right.
\end{eqnarray}
for the left-incident states and
\begin{eqnarray}
\label{ER}
    \psi_{E,R}(x)=C\left\{
    \begin{array}{l}
      t_R\alpha_{-q}^{(-)}e^{-iqx},\qquad  x\ll-\xi_d \medskip \\
      \alpha_{-q}^{(+)}e^{-iqx}+r_R\alpha_q^{(+)}e^{iqx},\qquad  x\gg\xi_d
    \end{array}\right.
\end{eqnarray}
for the right-incident states. Here $q(E)=\sqrt{E^2-\Delta_0^2}/|v_{F,x}|\geq 0$ and
$$
  \alpha^{(\pm)}_q=\frac{1}{\sqrt{2}}
    \left(\begin{array}{c}
    \dfrac{\Delta_\pm}{\Delta_0}\sqrt{1+\dfrac{v_{F,x}q}{E}}\,\mathrm{sign}\,E\\
    \sqrt{1-\dfrac{v_{F,x}q}{E}}
    \end{array}
    \right).
$$
The Andreev scattering states are normalized, in the $\delta$-function sense, with the normalization coefficient given by
\begin{equation}
\label{C-norm}
  C(E)=\frac{1}{\sqrt{2\pi|v_{F,x}|}}\sqrt{\frac{|E|}{\sqrt{E^2-\Delta_0^2}}}.
\end{equation}
The proof is presented in Appendix \ref{sec: normalization}.

The Andreev reflection and transmission coefficients $r_{L,R}(E)$ and $t_{L,R}(E)$ can be found by matching the asymptotics (\ref{EL}) and (\ref{ER}) and the solutions of Eq. (\ref{And-eq}) near the DW. 
They satisfy the following general properties, independent on the details of the DW structure: 
\begin{eqnarray}
\label{tr-props-1}
    &&|t_L|^2+|r_L|^2=|t_R|^2+|r_R|^2=1,\quad t_R^*r_L+r_R^*t_L=0,\\
\label{tr-props-2}
    &&\frac{t_L}{t_R}=\frac{\Delta_-}{\Delta_+}.
\end{eqnarray}
It follows from Eqs. (\ref{tr-props-1}) and (\ref{tr-props-2}) that $|t_L|=|t_R|=t$ and $|r_L|=|r_R|=r=\sqrt{1-t^2}$, and also that the scattering matrix defined as
\begin{equation}
\label{S-matrix}
    S=
    \left(\begin{array}{cc}
        t_L & r_R\\
        r_L & t_R
    \end{array}\right)
\end{equation}
is unitary.

Expressions (\ref{tr-props-1}) and (\ref{tr-props-2}) follow from certain ``conservation laws'' for the Andreev equation. Let $\psi_{E,p_1}=(u_{E,p_1},v_{E,p_1})^T$ and $\psi_{E,p_2}=(u_{E,p_2},v_{E,p_2})^T$
be two solutions of Eq. (\ref{And-eq}) corresponding to the same energy, with $p_{1,2}=L$ or $R$. We define an analog of the Wronskian as follows: $w[\psi_{E,p_1},\psi_{E,p_2}]=\tr(\psi_{E,p_1}^\dagger\hat\sigma_3\psi_{E,p_2})$.
It is straightforward to show that $dw/dx=0$, i.e. $w[\psi_{E,p_1},\psi_{E,p_2}]$ does not depend on $x$. Also, $\tilde\psi=\hat\sigma_1\psi^*$ corresponds to the same energy as $\psi$ and $w[\tilde\psi_{E,p_1},\psi_{E,p_2}]$ does not depend on $x$ either.
Therefore,
\begin{eqnarray}
\label{w12}
  &&u_{E,p_1}^*(x)u_{E,p_2}(x)-v_{E,p_1}^*(x)v_{E,p_2}(x)=\mathrm{const},\\
\label{wt12}
  &&v_{E,p_1}(x)u_{E,p_2}(x)-u_{E,p_1}(x)v_{E,p_2}(x)=\mathrm{const}.
\end{eqnarray}
The constants on the right-hand side can be calculated far from the DW, using the asymptotic expressions (\ref{EL}) and (\ref{ER}). 
The properties (\ref{tr-props-1}) and  (\ref{tr-props-2}) are obtained from Eqs. (\ref{w12}) and (\ref{wt12}), respectively. 

Explicit analytical expressions for the reflection and transmission coefficients can only be derived in some simple cases, in particular, for a ``sharp DW'' model, in which the DW thickness is sent to zero, see Appendix \ref{sec: sharp DW}. 
These expressions can be used to find the asymptotics of $t_{L,R}$ and $r_{L,R}$ for an arbitrary DW of a finite thickness $\xi_d$ at the energies close to the bulk gap edge, when the wavelength of the Andreev modes is much greater than $\xi_d$. 
It follows from Eq. (\ref{tr-sharp-DW}) that 
\begin{equation}
\label{tr-band edge}
    t_{L,R}=0,\quad
    r_{L,R}=-1,
\end{equation}
at $|E|=\Delta_0$. 

It is also possible to find the asymptotics of the reflection and transmission coefficients for an arbitrary DW at $|E|\gg\Delta_0$. At large energies, one can neglect the off-diagonal terms in the Andreev equations (\ref{And-eq}). 
Then, the solutions for the left- and right-incident modes have the form 
$$
  \psi_{E,L}(x)=\frac{1}{\sqrt{2\pi|v_{F,x}|}}\left(\begin{array}{c}
                         1\\0
                         \end{array}\right)e^{iEx/v_{F,x}},\quad
  \psi_{E,R}(x)=\frac{1}{\sqrt{2\pi|v_{F,x}|}}\left(\begin{array}{c}
                         0\\1
                         \end{array}\right)e^{-iEx/v_{F,x}},
$$
if $E/v_{F,x}>0$, and
$$
  \psi_{E,L}(x)=\frac{1}{\sqrt{2\pi|v_{F,x}|}}\left(\begin{array}{c}
                         0\\1
                         \end{array}\right)e^{-iEx/v_{F,x}},\quad
  \psi_{E,R}(x)=\frac{1}{\sqrt{2\pi|v_{F,x}|}}\left(\begin{array}{c}
                         1\\0
                         \end{array}\right)e^{iEx/v_{F,x}},
$$
if $E/v_{F,x}<0$. Comparing these expressions with Eqs. (\ref{EL}) and (\ref{ER}), we obtain:
\begin{equation}
\label{tr-large E}
  \begin{array}{l}
  t_L=\dfrac{\Delta_-}{\Delta_+},\ t_R=1,\ r_L=r_R=0,\quad \mathrm{for}\ E/v_{F,x}>0,\medskip\\
  t_L=1,\ t_R=\dfrac{\Delta_+}{\Delta_-},\ r_L=r_R=0,\quad \mathrm{for}\ E/v_{F,x}<0,
  \end{array}
\end{equation}
at $|E|\gg\Delta_0$.

\subsection{Bound states}
\label{sec: bound states}

At subgap energies, i.e. at $|E|<\Delta_0$, quasiparticles cannot propagate in the bulk, but Eq. (\ref{And-eq}) has solutions which are localized near the DW. It turns out that the number of such solutions 
can be expressed in terms of the properties of the scattering states:
\begin{equation}
\label{Nb-final}
  N_B=1+\frac{i}{2\pi}\ln\frac{\det S(\infty)\det S(-\infty)}{\det S(\Delta_0)\det S(-\Delta_0)},
\end{equation}
where $S(E)$ is the scattering matrix defined by Eq. (\ref{S-matrix}). The proof is presented in Appendix \ref{sec: proof of LT}. The expression (\ref{Nb-final}) plays the role of Levinson's theorem for the Andreev equation (recall that Levinson's theorem 
relates the number of bound states of the Schr\"odinger equation to the phase shifts of the scattering states, see Ref. \onlinecite{Newton-book}). Note that there is a 
formal similarity between the Andreev Hamiltonian for a superconducting DW and the Dirac Hamiltonian in one dimension. 
The analogs of Levinson's theorem for the Dirac equation have been extensively studied, in particular, in the context of soliton charge fractionalization.\cite{Dirac-1D}

The determinant of the $S$-matrix is a pure phase, which allows one to write the second term on the right-hand side of Eq. (\ref{Nb-final}) in a more transparent form. From the asymptotics (\ref{tr-band edge}) and (\ref{tr-large E}),  
we have $\det S(\pm\Delta_0)=-1$ and $\det S(\pm\infty)=e^{\pm i\zeta}$, where $\zeta=(2\theta-\chi)\sign v_{F,x}$. Therefore, $\ln[\det S(\pm\Delta_0)/\det S(\pm\infty)]=\pm i\pi\mp i\zeta+2i\pi n_\pm$ and 
\begin{equation}
\label{NB-winding}
  N_B=1+n_++n_-,
\end{equation}
where the integers $n_\pm$ are the winding numbers picked up by the phase of $\det S(E)$ as the energy varies between the bulk gap edges and the infinities. 
In particular, for the sharp DW model the determinant of the $S$-matrix is given by Eq. (\ref{det S-sharp DW}), from which we obtain $n_+=n_-=0$, therefore, $N_B=1$, in agreement with the direct calculation of the bound states 
in Appendix \ref{sec: sharp DW}.
Different models for the DW structure yield different results for the ABS spectrum: while $N_B=1$ for a sharp DW, one can have $N_B>1$ for a DW of finite width.\cite{Ho84}

At given $\bk_F=k_F(\cos\theta,\sin\theta)$, the bound states have the energies $E_a$ ($a=0,...,N_B-1$) and are asymptotically described by the wave functions $\psi_a(x)\sim e^{-\sqrt{\Delta_0^2-E_a^2}|x|/|v_{F,x}|}$, at $|x|\gg\xi_d$. 
As the angle $\theta$ varies between $0$ and $2\pi$, the energies change, forming the ABS bands $E_a(\theta)$. The BdG electron-hole symmetry is manifested in the following property:
\begin{equation}
\label{BdG-symmetry}
  E_a(\theta)=-E_a(\theta+\pi).
\end{equation}
We use the index $a=0$ to label the branch of the ABS's whose energy vanishes at some $\theta$ (zero modes). Note that if $E_0(\theta)=0$ then, according to Eq. (\ref{BdG-symmetry}), $E_0(\theta+\pi)=0$ as well. The existence of zero modes is dictated
by a topological argument, which relates the number of such modes with the difference between the $\bk$-space topological invariants of the chiral order parameters in the two domains.\cite{Vol92}
Taking the spin into account, there are two pairs of spin-degenerate zero modes corresponding to the opposite directions of semiclassical propagation, i.e. four zero modes altogether. This is confirmed by the explicit calculation for a sharp DW model in 
Appendix \ref{sec: sharp DW}. In general, fermion zero modes are present on any interface separating two superfluid or superconducting states with different topological charges, see Ref. \onlinecite{Vol-book} for a review.

An important property of the ABS bands is the absence of degeneracies: 
\begin{equation}
\label{no-degen}
  E_a(\theta)\neq E_b(\theta),
\end{equation}
at any $a$ and $b$. This can be shown as follows. Suppose that at some $\theta$ there are two bound-state solutions of Eq. (\ref{And-eq}), $\psi_1$ and $\psi_2$, corresponding to the same energy $E$. 
One can use Eq. (\ref{wt12}) with the constant on the right-hand side equal to zero (due to the exponential decay of the bound states at infinity) and obtain $\psi_1(x)=F(x)\psi_2(x)$, where $F(x)$ is a scalar function. 
Inserting this into Eq. (\ref{And-eq}), we have
$v_{F,x}\nabla_xF=0$, therefore $F(x)=\mathrm{const}$, i.e. $\psi_1$ and $\psi_2$ in fact describe the same state. The ABS's can only become degenerate when $v_{F,x}=0$, but
at such directions of $\bk_F$ the Andreev approximation is not applicable,

The bound states are responsible for a nonzero density of states at $|E|<\Delta_0$, which affects the system's low-temperature thermodynamics (for instance, there is a linear in $T$ contribution to the
specific heat, whose magnitude is proportional to the volume fraction occupied by the DWs, see Ref. \onlinecite{BK88}), and also influence the Josephson current between two chiral superconductors.\cite{ABS-Josephson}

\section{Domain wall friction and mass}
\label{sec: M and eta}

Now we are in the position to calculate ${\cal N}(\epsilon,\epsilon')$ in the Gaussian kernel (\ref{K-gen-int}). In the semiclassical approximation, the states $i$ and $j$ in Eq. (\ref{cal N def}) 
correspond to the same Fermi wavevector $\bk_F$, so that the matrix elements can be taken between the solutions of the Andreev equation (\ref{And-eq}). We have
\begin{eqnarray}
\label{cal N}
    {\cal N}(\epsilon,\epsilon')=\frac{1}{(\epsilon-\epsilon')^2}\bigl[\varphi_1(\epsilon,\epsilon')\theta(|\epsilon|-\Delta_0)\theta(|\epsilon'|-\Delta_0)\nonumber\\
    +\varphi_2(\epsilon,\epsilon')\theta(|\epsilon|-\Delta_0)\theta(\Delta_0-|\epsilon'|)\nonumber\\
    +\varphi_2(\epsilon',\epsilon)\theta(\Delta_0-|\epsilon|)\theta(|\epsilon'|-\Delta_0)\nonumber\\
    +\varphi_3(\epsilon,\epsilon')\theta(\Delta_0-|\epsilon|)\theta(\Delta_0-|\epsilon'|)\bigr].
\end{eqnarray}
At given $\bk_F$, ${\cal N}(\epsilon,\epsilon')$ contains contributions from the transitions between different scattering states ($\varphi_1$), between the scattering states and the bound states ($\varphi_2$), 
and also between different bound states ($\varphi_3$). 
The total intensity of the quasiparticle transitions is obtained by summing over all directions of semiclassical propagation and over the two spin projections:
\begin{eqnarray}
\label{nus-defs}
  &&\varphi_1(\epsilon,\epsilon')=2\sum_{\bk_F}\sum_{p,p'}|\langle\epsilon,p|\hat Q|\epsilon',p'\rangle|^2,\nonumber\\
  &&\varphi_2(\epsilon,\epsilon')=2\sum_{\bk_F}\sum_{p}\sum_b|\langle\epsilon,p|\hat Q|b\rangle|^2\delta(\epsilon'-E_b),\\
  &&\varphi_3(\epsilon,\epsilon')=2\sum_{\bk_F}\sum_{a\neq b}|\langle a|\hat Q|b\rangle|^2\delta(\epsilon-E_a)\delta(\epsilon'-E_b).\nonumber
\end{eqnarray}
Here 
$$
  \hat Q=\left(\begin{array}{cc}
     0 & \dfrac{d\Delta_{\bk_F}(x)}{dx} \\
     \dfrac{d\Delta^*_{\bk_F}(x)}{dx} & 0
  \end{array}\right),
$$
with $\Delta_{\bk_F}(x)$ given by Eq. (\ref{Andreev-OP}), $p,p'=L$ or $R$, and $a,b=0,...,N_B-1$. The summation over $\bk_F$ amounts to the integration over the angle $\theta$:
$$
  \sum_{\bk_F}(...)=N_F\int_0^{2\pi}d\theta|v_{F,x}|(...),
$$
where $N_F=m^*/2\pi$ is the Fermi-level density of states. It follows from the electron-hole symmetry of the quasiparticle spectrum that ${\cal N}(-\epsilon,-\epsilon')={\cal N}(\epsilon,\epsilon')$ and, therefore, 
$\varphi_i(-\epsilon,-\epsilon')=\varphi_i(\epsilon,\epsilon')$.

To calculate the matrix elements in Eq. (\ref{nus-defs}) one needs to know the wave functions of the bound and scattering states. Lacking such knowledge for a general DW, one can still make progress using dimensional arguments. Since
the only energy and length scales of the problem are given by $\Delta_0$ and $\xi_d$ respectively, it is not difficult to show that
\begin{eqnarray}
\label{F_123}
  &&\sum_{p,p'}|\langle\epsilon,p|\hat Q|\epsilon',p'\rangle|^2=\frac{\Delta_0^2}{v_{F,x}^2}F_1\left(\frac{\epsilon}{\Delta_0},\frac{\epsilon'}{\Delta_0};\theta\right),\nonumber\\
  &&\sum_{p}\sum_b|\langle\epsilon,p|\hat Q|b\rangle|^2\delta(\epsilon'-E_b)=\frac{\Delta_0^2}{v_{F,x}^2}F_2\left(\frac{\epsilon}{\Delta_0},\frac{\epsilon'}{\Delta_0};\theta\right),\\
  &&\sum_{a\neq b}|\langle a|\hat Q|b\rangle|^2\delta(\epsilon-E_a)\delta(\epsilon'-E_b)=\frac{\Delta_0^2}{v_{F,x}^2}F_3\left(\frac{\epsilon}{\Delta_0},\frac{\epsilon'}{\Delta_0};\theta\right),\nonumber
\end{eqnarray}
where $F_i(x,x';\theta)$ are dimensionless functions. Inserting these expressions in Eq. (\ref{nus-defs}), we obtain: 
\begin{eqnarray}
\label{nus-xx-1}
  \varphi_i(\epsilon,\epsilon')=\frac{2N_F\Delta_0^2}{v_F}\int_0^{2\pi}\frac{d\theta}{|\cos\theta|}F_i\left(\frac{\epsilon}{\Delta_0},\frac{\epsilon'}{\Delta_0};\theta\right)\nonumber\\
  =\frac{N_F\Delta_0^2}{v_F}f_i\left(\frac{\epsilon}{\Delta_0},\frac{\epsilon'}{\Delta_0}\right),
\end{eqnarray}
where $f_i$ are dimensionless functions. The Fermi surface angular integrals logarithmically diverge at $\theta\to\pm\pi/2$ (i.e. for the quasiparticles moving almost parallel to the DW) and 
have to be cut off at $|\cos\theta|\sim\sqrt{\Delta_0/\epsilon_F}\ll 1$. In this way we obtain, with logarithmic accuracy: $f_i=\ln(\epsilon_F/\Delta_0)\tilde f_i$, where $\epsilon_F=k_F^2/2m$ is the Fermi energy and
$\tilde f_i(x,x')=2F_i(x,x';\pi/2)+2F_i(x,x';-\pi/2)$. Therefore,
\begin{equation}
\label{nus-xx-2} 
  \varphi_i(\epsilon,\epsilon')=\frac{N_F\Delta_0^2}{v_F}\ln\left(\frac{\epsilon_F}{\Delta_0}\right)\tilde f_i\left(\frac{\epsilon}{\Delta_0},\frac{\epsilon'}{\Delta_0}\right).
\end{equation}

The expressions above can be calculated explicitly in the case of a sharp DW.\cite{Sam11} It follows from Eq. (\ref{sharp-DW-model}) that
$$
  \hat Q=\Delta_0\left(\begin{array}{cc}
     0 & \rho \\
     \rho^* & 0
  \end{array}\right)\delta(x),
$$
where $\rho=(\Delta_+-\Delta_-)/\Delta_0$. Using Eqs. (\ref{EL}) and (\ref{ER}), we obtain:
\begin{equation}
\label{F1}
  F_1(x,x';\theta)=\frac{2}{\pi^2}\sin^2\left(\theta-\frac{\chi}{2}\right)\frac{|xx'|t^2(x)t^2(x')}{\sqrt{x^2-1}\sqrt{x^{\prime,2}-1}}\left[1-\frac{1}{xx'}\cos^2\left(\theta-\frac{\chi}{2}\right)\right].
\end{equation}
Here $t$ is the absolute value of the Andreev transmission coefficient, see Eq. (\ref{tr-sharp-DW}), for which we have
$$
  t^2(x)=\frac{x^2-1}{x^2-\cos^2(\theta-\chi/2)}.
$$
Using Eqs. (\ref{EL}), (\ref{ER}), and (\ref{psi-bound}), we obtain:
\begin{eqnarray}
\label{F2}
  F_2(x,x';\theta)&=&\frac{2}{\pi}\sin^2\left(\theta-\frac{\chi}{2}\right)\frac{|x|t^2(x)}{\sqrt{x^2-1}}\sqrt{1-x^{\prime,2}}\nonumber\\
  &&\times\Bigl[1-\frac{x'}{x}\cos(2\theta-\chi)-\frac{\sqrt{1-x^{\prime,2}}}{x}\sign(\cos\theta)\sin(2\theta-\chi)\Bigr]\delta\left[x'-\frac{E_0(\theta)}{\Delta_0}\right],
\end{eqnarray}
where the bound state energy $E_0$ is given by Eq. (\ref{E-bound}). Finally, $F_3=0$, since there is only one bound state at each $\theta$. 
One can see that $F_i$ are nonsingular near the bulk gap edge, i.e. at $|x|,|x'|\to 1$, because of the vanishing of the Andreev transmission coefficients. This property actually holds for a general DW, since 
the wavelength of the Andreev states near the bulk gap edge diverges, making it possible to neglect the DW width at $|E|\to\Delta_0$. 

Eqs. (\ref{F1}) and (\ref{F2}) yield the following expressions for the energy dependence in Eq. (\ref{nus-xx-2}):
\begin{eqnarray}
\label{tilde-f1}
  &&\tilde f_1(x,x')=\frac{8}{\pi^2}\cos^2\frac{\chi}{2}\frac{|xx'|\sqrt{x^2-1}\sqrt{x^{\prime,2}-1}}{[x^2-\sin^2(\chi/2)][x^{\prime,2}-\sin^2(\chi/2)]}\Bigl(1-\frac{1}{xx'}\sin^2\frac{\chi}{2}\Bigr),\\
\label{tilde-f2}
  &&\tilde f_2(x,x')=\frac{4}{\pi}\cos^3\frac{\chi}{2}\frac{|x|\sqrt{x^2-1}}{x^2-\sin^2(\chi/2)}\Bigl(1-\frac{x'}{x}\Bigr)\left[\delta\left(x'-\sin\frac{\chi}{2}\right)+\delta\left(x'+\sin\frac{\chi}{2}\right)\right],\\
\label{tilde-f3}
  &&\tilde f_3(x,x')=0.
\end{eqnarray}
In some exceptional cases the logarithmic approximation, see Eq. (\ref{nus-xx-2}), might be insufficient. For instance, it follows from Eqs. (\ref{tilde-f1})-(\ref{tilde-f3}) that the functions $\tilde f_i$ all
vanish for a sharp DW with $\chi=\pi$. In such cases one should use the more general Eq. (\ref{nus-xx-1}).

\subsection{Results}
\label{sec: Results}

Now we turn to the calculation of the effective dynamic action for the DW, see Eq. (\ref{K-gen-int}). At finite temperatures, the most singular contribution to the action at $\nu_m\to 0$ comes from $\epsilon'$ close to $\epsilon$, 
which means that one can keep only the first term in the expression (\ref{cal N}). The contribution from the transitions between the bound states can be neglected, because the ABS bands are non-degenerate, see Eq. (\ref{no-degen}). We have
$$
  {\cal K}(\nu_m)=\frac{\nu_m^2}{2}\int_{\Delta_0}^\infty d\epsilon\,\left(-\frac{\partial f}{\partial\epsilon}\right)\varphi_1(\epsilon,\epsilon)
  \int_{-\infty}^{\epsilon-\Delta_0}\frac{d\varepsilon}{\varepsilon^2+\nu_m^2},
$$
where $\varepsilon=\epsilon-\epsilon'$. The integration over $\varepsilon$ can be extended to infinity, since we are only interested in the low-frequency limit. 
Inserting here $\varphi_1$ given by Eq. (\ref{nus-xx-2}), we obtain that the leading frequency dependence of the kernel (\ref{K-gen}) is non-analytic: ${\cal K}(\nu_m)=\eta|\nu_m|$, with the viscous friction coefficient given by
\begin{equation}
\label{eta}
  \eta(T)=\frac{N_F\Delta_0^2}{v_F}\ln\left(\frac{\epsilon_F}{\Delta_0}\right)\Phi\left(\frac{\Delta_0}{T}\right),
\end{equation}
where
$$
  \Phi(y)=\frac{\pi y}{8}\int_1^\infty\frac{dx}{\cosh^2(yx/2)}\tilde f_1(x,x).
$$
It is not possible to calculate the integral and obtain an analytical expression for $\Phi(y)$. Even in the simplest case of a sharp DW, it follows from Eq. (\ref{tilde-f1}) that 
$$
  \Phi(y)=\frac{y}{\pi}\cos^2\frac{\chi}{2} \int_1^\infty\frac{dx}{\cosh^2(yx/2)}\frac{x^2-1}{x^2-\sin^2(\chi/2)},
$$
which still cannot be evaluated in a closed form.

At low temperatures, all we need is the asymptotics of $\Phi(y)$ at $y\gg 1$, which is given by $\Phi(y)\sim e^{-y}$. Therefore,
$$
  \eta(T)\sim\frac{N_F\Delta_0^2}{v_F}\ln\left(\frac{\epsilon_F}{\Delta_0}\right)e^{-\Delta_0/T}, 
$$
at $T\ll\Delta_0$. Physically, the DW friction is caused by the transitions between the bulk scattering states. These states absorb energy from the DW and then carry it away to dissipate into the thermal reservoir. Since 
the bulk quasiparticles are gapped, the temperature dependence of the friction coefficient is exponential. 

Note that the expression (\ref{eta}) has the same general order of magnitude, but a different temperature dependence, as the friction coefficient of an $A-B$ interface in superfluid
${}^3$He (Refs. \onlinecite{AB-boundary-1} and \onlinecite{AB-boundary-2}). In the latter case, there are low-energy quasiparticles in the bulk with the momenta close to the $A$-phase gap nodes. The interface friction at low temperatures is 
dominated by the Andreev reflection of such quasiparticles off the interface, leading to a power-law behaviour $\eta(T)\propto T^4$ or $T^3$, depending on the orientation of the orbital vector $\bm{l}$ relative to the interface.\cite{AB-boundary-2}  

At $T\to 0$, the friction is negligibly small and the DW dynamics is dominated by inertia. Setting $\nu_m=0$ inside the integral in Eq. (\ref{K-gen-int}), we obtain ${\cal K}(\nu_m)=M\nu_m^2/2$, where
\begin{equation}
\label{M-gen}
  M=\int_0^\infty d\epsilon\int_{-\infty}^0d\epsilon'\,\frac{{\cal N}(\epsilon,\epsilon')}{\epsilon-\epsilon'}
\end{equation}
is the effective mass per unit length of the DW. Inserting here Eqs. (\ref{cal N}) and (\ref{nus-xx-2}), we obtain
\begin{equation}
\label{mass}
  M=C\frac{N_F\Delta_0}{v_F}\ln\left(\frac{\epsilon_F}{\Delta_0}\right),
\end{equation}
where
\begin{equation}
\label{mass-C-def}
  C=\int_1^\infty dx\int_{-\infty}^{-1}dx'\,\frac{\tilde f_1(x,x')}{(x-x')^3}+2\int_1^\infty dx\int_{-1}^0dx'\,\frac{\tilde f_2(x,x')}{(x-x')^3}+\int_0^1 dx\int_{-1}^0 dx'\,\frac{\tilde f_3(x,x')}{(x-x')^3}
\end{equation}
is a dimensionless coefficient. We see that the transitions between the electron and hole branches of the continuous spectrum (the first term on the right-hand side), between the bound and the scattering states (the second term), and 
between different bound states (the third term) all contribute to the DW inertial mass. 

Since $N_F=m^*/2\pi$ in two dimensions, we have $M\sim(m^*/\xi_0)\ln(\epsilon_F/\Delta_0)$, where $\xi_0\sim v_F/\Delta_0$ is the pair coherence length. 
At zero temperature $\xi_d\sim\xi_0$, therefore, the ratio of the DW mass to the total fermionic mass per unit length contained in a strip of width $\xi_d$ is of the order of $(\Delta_0/\epsilon_F)^2\ln(\epsilon_F/\Delta_0)$, i.e. very small.
This is consistent with the estimate of the effective mass of the $A-B$ interface: according to Ref. \onlinecite{AB-boundary-1}, it is smaller by a factor of $(\Delta_0/\epsilon_F)^2$ than the total mass of the superfluid in the interface region.

\section{Conclusions}

We have calculated the viscous friction coefficient and the zero-temperature effective mass of a domain wall in a chiral $p$-wave superconductor. The origin of both can be traced to the interactions of the DW with 
the Bogoliubov fermionic quasiparticles. The viscous friction is caused by the transitions among the bulk quasiparticle states induced by the DW motion. The friction coefficient is exponentially small at low temperatures, due to 
the bulk quasiparticles requiring thermal activation. The effective mass is determined by the transitions involving both the scattering states in the bulk and the Andreev bound states localized near the wall. 

The classical equation of motion for a DW can be written as $M\ddot X+\eta\dot X=F$, where the effective mass $M$ and the friction coefficient $\eta$ are given by Eqs. (\ref{mass}) and (\ref{eta}), respectively. The right-hand side contains the external driving force, which can come from the interaction 
of the orbital magnetization of the Cooper pairs with the external magnetic field,\cite{Legg77} or from the coupling of the superconducting order parameter with the lattice deformation created by a sound wave.\cite{SU91} The latter mechanism offers a direct way of measuring the
dynamical characteristics of the DW's by probing their contribution to the ultrasound attenuation.\cite{JRU86} 

The DW dynamics for small deviations from equilibrium are determined by the Bogoliubov quasiparticle spectrum in the presence of a static DW. We presented a detailed investigation of this spectrum, including 
the general properties of the scattering states as well as the derivation of an analog of Levinson's theorem counting the number of the Andreev bound states. 
We did most of our calculations for a general DW with an arbitrary phase difference between the domains, without relying on any particular model for a DW structure. 
To illustrate the general formulas, we discussed in detail the case of a DW of zero width, for which one can make considerable analytical progress. 

Our results are immediately applicable to clean neutral fermionic superfluids, such as ${}^3$He or cold atomic Fermi gases. In real superconductors, one has to take into account magnetic fields and screening currents, as well as disorder. 
In particular, impurities are expected to change the DW dynamics qualitatively, due to pinning.

\acknowledgments

This work was supported by a Discovery Grant from the Natural Sciences and Engineering Research Council of Canada.

\appendix

\section{Ginzburg-Landau description}
\label{sec: GL description}

To develop a phenomenological understanding of the static DW structure, in particular, the origin of the phase difference between the domains, we use the Ginzburg-Landau (GL) free energy density given by $F=F_u+F_g$, where
\begin{equation}
\label{GL-energy-u}
  F_u=\alpha(|\eta_1|^2+|\eta_2|^2)+\beta_1(|\eta_1|^2+|\eta_2|^2)^2+\beta_2|\eta_1^2+\eta_2^2|^2
\end{equation}
is the uniform part and
\begin{equation}
\label{GL-energy-g}
  F_g=K_1(\nabla_i\eta_j)^*(\nabla_i\eta_j)+K_2(\nabla_i\eta_i)^*(\nabla_j\eta_j)+K_3(\nabla_i\eta_j)^*(\nabla_j\eta_i)
\end{equation}
is the gradient part. To avoid unnecessary complications we use the expression appropriate for the isotropic case. 
The chiral states $\bmeta=\Delta_0(1,\pm i)$, with $\Delta_0=\sqrt{|\alpha|/4\beta_1}$ correspond to the minimum of $F_u$ if $\beta_1,\beta_2>0$.

The DW structure, see Eq. (\ref{DW-general-structure}), can be written as follows:
\begin{equation}
\label{DW-x}
  \eta_1(x)=\Delta_0f_1(x)e^{i\phi(x)},\quad \eta_2(x)=\Delta_0f_2(x)e^{i\phi(x)-i\gamma(x)},
\end{equation}
where $f_{1,2}$ are dimensionless amplitudes of the order parameter components. The order parameter asymptotics are given by Eq. (\ref{DW-model}).

The origin of a nonzero phase difference $\chi$ can be traced to the condition of vanishing supercurrent across the DW. The latter is obtained from Eq. (\ref{GL-energy-g}) in the standard manner, with the result
$j_i=2\im\,(K_1\eta_j^*\nabla_i\eta_j+K_2\eta_i^*\nabla_j\eta_j+K_3\eta_j^*\nabla_j\eta_i)$. Inserting here Eq. (\ref{DW-x}), we obtain:
\begin{equation}
\label{j_x}
  j_x=2\Delta_0^2(K_{123}f_1^2+K_1f_2^2)(\nabla_x\phi)-2K_1\Delta_0^2f_2^2(\nabla_x\gamma)
\end{equation}
where $K_{123}=K_1+K_2+K_3$. The presence in this expression of both the common and the relative phase gradients reflects the intimate coupling of the gauge and the internal (orbital) degrees of freedom in $p$-wave fermionic superfluids.

Because of the current conservation we have $\nabla_xj_x=0$, therefore $j_x=\mathrm{const}$. The value of the transverse current is fixed by external sources and one can set $j_x=0$ at all $x$. Then, Eq. (\ref{j_x}) yields a linear
relation between the gradients of $\phi$ and $\gamma$, which allows one to eliminate the common phase from the GL energy functional. The result is as follows:
\begin{equation}
\label{F-ug-reduced}
  \begin{array}{l}
  F_u=\alpha\Delta_0^2(f_1^2+f_2^2)+\beta_1\Delta_0^4(f_1^2+f_2^2)^2+\beta_2\Delta_0^4(f_1^4+f_2^4+2f_1^2f_2^2\cos 2\gamma),\\ \\
  F_g=K_{123}\Delta_0^2(\nabla_xf_1)^2+K_1\Delta_0^2(\nabla_xf_2)^2+\dfrac{K_1K_{123}f_1^2f_2^2}{K_{123}f_1^2+K_1f_2^2}\Delta_0^2(\nabla_x\gamma)^2.
  \end{array}
\end{equation}
Variational minimization of these expressions yields a system of three coupled nonlinear differential equations for $f_{1,2}(x)$ and $\gamma(x)$,
subject to the boundary conditions $f_{1,2}(\pm\infty)=1$ and $\gamma(\pm\infty)=\pm\pi/2$. 
Using the solution of these equations, we can calculate the parameter $\chi$ in Eq. (\ref{DW-model}):
\begin{equation}
\label{phi-gamma}
  \chi\equiv\phi(+\infty)-\phi(-\infty)=\int_{-\infty}^\infty\frac{K_1f_2^2}{K_{123}f_1^2+K_1f_2^2}\frac{d\gamma}{dx}\,dx.
\end{equation}
The value of $\chi$ is manifestly non-universal, in the sense that it depends on the microscopic details. 
We note that, while the ``locking'' between $\phi$ and $\gamma$, which results in the relation (\ref{phi-gamma}), is due to the condition $j_x=0$, the supercurrent along the DW remains nonzero.

Due to the complexity of the equations for $f_{1,2}(x)$ and $\gamma(x)$, there is no exact analytical solution for the DW structure. To make progress, one can use, e.g. a constant-amplitude ansatz for the order parameter components,\cite{VG85}
which amounts to putting $f_{1,2}(x)=1$ at all $x$. Then, we obtain from Eq. (\ref{F-ug-reduced}) the following expression for the free energy: 
$$
  F=(...)+\tilde K\Delta_0^2(\nabla_x\gamma)^2+2\beta_2\Delta_0^4\cos 2\gamma,
$$
where the first term contains the $\gamma$-independent contributions and $\tilde K=K_1K_{123}/(K_{123}+K_1)$. The variational equation for the relative phase has the form of a sine-Gordon equation, with a kink-like solution $\sin\gamma(x)=\tanh(x/\xi_d)$, where
$\xi_d=\sqrt{\tilde K/4\beta_2\Delta_0^2}$ is of the order of the GL correlation length and has the meaning of the DW thickness. From Eq. (\ref{phi-gamma}) we have
\begin{equation}
\label{phi-gamma-VG}
  \chi=\frac{K_1}{2K_1+K_2+K_3}\pi.
\end{equation}
In the weak coupling model, $K_1=K_2=K_3$ (Ref. \onlinecite{Book}), therefore, $\chi=\pi/4$.

\section{Normalization of the Andreev scattering states}
\label{sec: normalization}

Let us consider two solutions, $\psi_{E,p}$ and $\psi_{E',p'}$, of Eq. (\ref{And-eq}), corresponding to energies $E$ and $E'$, with $p,p'=L$ or $R$.
It is easy to show that $iv_{F,x}\nabla_x\tr(\psi_{E,p}^\dagger\hat\sigma_3\psi_{E',p'})=(E-E')\tr(\psi_{E,p}^\dagger\psi_{E',p'})$,
where ``$\tr$'' denotes a $2\times 2$ matrix trace in the electron-hole space (setting $E=E'$, we recover the ``conservation law'' for $w[\psi_{E,p},\psi_{E,p'}]$, see Sec. \ref{sec: scattering states}). After integration, 
we arrive at the following useful identity:
\begin{equation}
\label{psi12-identity}
    (E-E')\int_{x_1}^{x_2}\tr(\psi_{E,p}^\dagger\psi_{E',p'})\,dx=
        iv_{F,x}\left.\tr(\psi_{E,p}^\dagger\hat\sigma_3\psi_{E',p'})\right|_{x_1}^{x_2},
\end{equation}
which is valid for arbitrary $x_1$ and $x_2$. 

The next step is to put $x_1=-\ell/2$, $x_2=\ell/2$, and take the limit $\ell\to\infty$. The integral on the
left-hand side of Eq. (\ref{psi12-identity}) becomes the inner product of the states $\psi_{E,p}$ and
$\psi_{E',p'}$, denoted by $\langle E,p|E',p'\rangle$. To prove the normalization, it is sufficient to consider the case $E'\to E$. 
The normalization integral for the scattering states of the Schr\"odinger equation contains a Dirac $\delta$-function,\cite{LL3} i.e. should be interpreted as a generalized function, 
and we expect the same to hold for the Andreev scattering states as well. After the substitution of the asymptotic expressions (\ref{EL}) and (\ref{ER}), 
the right-hand side of Eq. (\ref{psi12-identity}) contains the terms proportional to $e^{\pm i(q+q')\ell/2}$ and $e^{\pm i(q-q')\ell/2}$, where $q'=q(E')$. 
The former terms oscillate fast at $\ell\to\infty$ and can be neglected.

Since the terms containing $e^{\pm i(q-q')\ell/2}$ oscillate fast unless $q=q'$, one can put $E=E'$ in the pre-exponential coefficient. Then, using the properties (\ref{tr-props-1}), we obtain
\begin{eqnarray}
\label{RRLL}
  &&\langle E,L|E',L\rangle=\langle E,R|E',R\rangle=\frac{2v_{F,x}^2q}{E}C^2(E)\lim_{\ell\to\infty}\frac{\sin(q-q')\ell/2}{E-E'},\\
  &&\langle E,L|E',R\rangle=\langle E,R|E',L\rangle=0.\nonumber
\end{eqnarray}
The identity $\lim_{\ell\to\infty}\sin(x\ell)/x=\pi\delta(x)$ allows one to write the last factor on the right-hand side of Eq. (\ref{RRLL}) in the form 
$$
    \lim_{\ell\to\infty}\frac{\sin(q-q')\ell/2}{E-E'}=\pi\sign E\,\delta(E-E').
$$
Substituting in Eq. (\ref{RRLL}) the expressions (\ref{C-norm}) for the normalization coefficients, we finally obtain:
$\langle E,p|E',p'\rangle=\delta_{pp'}\delta(E-E')$.

\section{``Sharp DW'' model}
\label{sec: sharp DW}

Many qualitative features of the DW quasiparticle spectrum can be illustrated using a simple model, in which there is a sharp boundary at $x=0$ between the two domains with uniform order parameters of opposite chirality.
The order parameter in the Andreev equation (\ref{And-eq}) has the form
\begin{equation}
\label{sharp-DW-model}
    \Delta_{\bk_F}(x)=\Delta_-\theta(-x)+\Delta_+\theta(x),
\end{equation}
where $\Delta_\pm$ are given by Eq. (\ref{Delta-pm}) and $\theta(x)$ is the Heaviside step function. The boundary condition for the Andreev wave functions is $\psi(+0)=\psi(-0)$.

For the scattering states ($|E|\geq\Delta_0$), one can use Eqs. (\ref{EL}) and (\ref{ER}) at all $x$. A straightforward calculation produces the following expressions 
for the transmission and reflection coefficients,
\begin{equation}
\label{tr-sharp-DW}
      \begin{array}{l}
      t_L=\dfrac{\Delta_-}{\Delta_+}t_R=\dfrac{2\Delta_-v_{F,x}q}{(\Delta_++\Delta_-)v_{F,x}q+(\Delta_+-\Delta_-)E},\medskip\\
      r_L=r_R=-\dfrac{\Delta_0(\Delta_+-\Delta_-)\sign E}{(\Delta_++\Delta_-)v_{F,x}q+(\Delta_+-\Delta_-)E},
      \end{array}
\end{equation}
and also for the determinant of the $S$-matrix, see Eq. (\ref{S-matrix}):
\begin{equation}
\label{det S-sharp DW}
  \det S(E)=\frac{\sqrt{E^2-\Delta_0^2}+i\lambda E}{\sqrt{E^2-\Delta_0^2}-i\lambda E},
\end{equation}
where $\lambda=\tan(\theta-\chi/2)\,\sign v_{F,x}$.

\begin{figure}
\includegraphics[width=8cm]{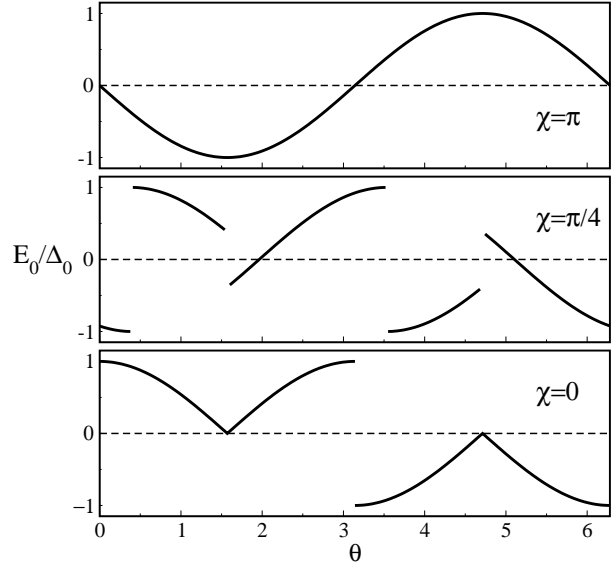}
\caption{The ABS energy for a sharp DW as a function of the direction of the quasiparticle propagation, for different values of the phase difference across the DW: 
$\chi=0$ (bottom panel), $\chi=\pi/4$ (middle panel), and $\chi=\pi$ (top panel).}
\label{fig: ABS-energy}
\end{figure}

For the subgap bound states ($|E|<\Delta_0$), the normalized wave function has the form
\begin{equation}
\label{psi-bound}
	\psi(x)=\sqrt{\frac{\Omega}{2|v_{F,x}|}}e^{-\Omega|x|/|v_{F,x}|}
	\left(\begin{array}{c}
	\dfrac{\Delta_\pm}{E\mp i\Omega\sign v_{F,x}} \\ 1
	\end{array}\right),
\end{equation}
where $\Omega=\sqrt{\Delta_0^2-E^2}$. The upper (lower) sign corresponds to $x>0$ ($x<0$). Matching the wave functions at $x=0$ we arrive at the characteristic equation
\begin{equation}
\label{E-eq-b}
       \frac{E+i\Omega\sign v_{F,x}}{E-i\Omega\sign v_{F,x}}=\frac{\Delta_-}{\Delta_+}.
\end{equation}
It has only one solution, which can be found as follows. Let us introduce $\tilde E=E\sign v_{F,x}$. Since $\tilde E^2+\Omega^2=\Delta_0^2$, one can write $\tilde E=\Delta_0\cos\Theta$, $\Omega=\Delta_0\sin\Theta$. It follows from Eq. (\ref{E-eq-b}) that $e^{2i\Theta}=e^{i(2\theta-\chi)}$,
therefore, $\Theta=\theta-\chi/2+\pi n$ and $\tilde E=\Delta_0(-1)^n\cos(\theta-\chi/2)$. The integer $n$ is found from the condition that $\Omega\geq 0$, which yields $\sign(\sin\Theta)=(-1)^n\sign\left[\sin\left(\theta-\chi/2\right)\right]=1$. 
Collecting all pieces together, we obtain the following expression for the ABS energy:
\begin{equation}
\label{E-bound}
      E_0(\theta)=\Delta_0s(\theta)\cos\left(\theta-\frac{\chi}{2}\right),
\end{equation}
where $s(\theta)=\sign\left[\sin\left(\theta-\chi/2\right)\cos\theta\right]$.

Expression (\ref{E-bound}) is valid for an arbitrary phase difference across the DW, thus generalizing the results of Refs. \onlinecite{Ho84,Naka86,BK88}, and \onlinecite{MS99}, in which 
the bound states were studied for the $(1,\pm i)$ (i.e. $\chi=0$) and $(\pm 1,i)$ (i.e. $\chi=\pi$) DWs. Note that the ABS energy is not a continuous function of $\theta$, in general. 
The discontinuities occur at the special directions of semiclassical propagation: at $\theta=\pm\pi/2$, i.e. for the quasiparticles moving parallel to the DW (in this case the Andreev approximation is actually not applicable and a more accurate
treatment is needed), and also at 
$\theta=\chi/2$ and $\theta=\chi/2+\pi$, for which the DW is ``invisible'' to the quasiparticles, because $\Delta_+=\Delta_-$. 

In Fig. \ref{fig: ABS-energy}, we plotted the ABS energy for several values of $\chi$. 
In particular, if $\chi=\pi$, then Eq. (\ref{E-bound}) yields $E=-\Delta_0\sin\theta=-\Delta_0k_y/k_F$ (see also Ref. \onlinecite{BK88}), vanishing at $k_y=0$. 
The presence of zero modes is in fact generic: the ABS energy vanishes at $\theta=(\chi\pm\pi)/2$, resulting in low-energy quasiparticles bound to the DW.
Taking the spin into account, we have two pairs of spin-degenerate zero mode branches.

We would like to note that one can also use Eq. (\ref{E-eq-b}) to obtain the ABS spectrum for a DW between two isotropic $s$-wave superconductors, which can be realized as a Josephson junction with the phase difference $\chi$.
In this case, the right-hand side of Eq. (\ref{E-eq-b}) is equal to $e^{-i\chi}$ and the bound state energy is given by $E=-\Delta_0\sign v_{F,x}\cos(\chi/2)$
(Ref. \onlinecite{s-wave-subgap}).

\section{Proof of Eq. (\ref{Nb-final})}
\label{sec: proof of LT}

We start with the Andreev Hamiltonian $H$ defined by Eq. (\ref{And-eq}), with the order parameter written in the form $\Delta_{\bk_F}=|\Delta_{\bk_F}|e^{i\varphi}$, where the amplitude and the phase 
have the following asymptotics: $\varphi(x)=\theta$ at $x\ll-\xi_d$, $\varphi(x)=\chi-\theta$ at $x\gg\xi_d$, and $|\Delta_{\bk_F}(x)|=\Delta_0$ at $|x|\gg\xi_d$. To represent the DW as a localized perturbation, we remove the phase 
from the off-diagonal elements by a unitary transformation as follows: $U^\dagger HU=\tilde H$, where $U=e^{i\varphi\hat\sigma_3/2}$. The transformed Hamiltonian is given by
$\tilde H=H_0+\delta H$, where $H_0=-iv_{F,x}\hat\sigma_0\nabla_x+\Delta_0\hat\sigma_1$ describes the Bogoliubov excitations in the uniform $p$-wave superconducting state, and
$\delta H=v_{F,x}\varphi'(x)\hat\sigma_0/2+(|\Delta_{\bk_F}(x)|-\Delta_0)\hat\sigma_1$ is a perturbation which is nonzero only near the DW. 

The number of the bound states for $H$ is the same as for $\tilde H$. To calculate the latter we observe that there is a one-to-one correspondence between the eigenstates of $\tilde H$ and $H_0$, 
which are found from the equations $\tilde H\tilde\psi_i=E_i\tilde\psi_i$ and $H_0\psi_i^{(0)}=E_i^{(0)}\psi_i^{(0)}$, respectively (to verify this one can introduce $\tilde H_\lambda=H_0+\lambda\delta H$ and consider 
the smooth evolution of the spectrum as the parameter $\lambda$ varies from $0$ to $1$). The total number of states is ``conserved'', which is formally expressed by the formula $\int_{-\infty}^\infty d\epsilon[\rho(\epsilon)-\rho_0(\epsilon)]=0$, where
\begin{equation}
\label{rho}
  \rho(\epsilon)=\lim_{\ell\to\infty}\int_{-\ell/2}^{\ell/2}dx\sum_i\delta(\epsilon-E_i)\tr\,\tilde\psi_i^\dagger(x)\tilde\psi_i(x)
\end{equation}
is the density of states for $\tilde H$ and
\begin{equation}
\label{rho_0}
  \rho_0(\epsilon)=\lim_{\ell\to\infty}\int_{-\ell/2}^{\ell/2}dx\sum_i\delta(\epsilon-E^{(0)}_i)\tr\,\psi_i^{(0),\dagger}(x)\psi_i^{(0)}(x)
\end{equation}
is the density of states for $H_0$. One can write $\rho=\rho_B+\rho_S$, where $\rho_B$ ($\rho_S$) is the contribution of the bound (scattering) states. On the other hand, $\rho_0$ is nonzero only at $|E|\geq\Delta_0$. Therefore,
$$
  0=\int_{|E|<\Delta_0}\rho_B\,d\epsilon+\int_{|E|\geq\Delta_0}(\rho_S-\rho_0)d\epsilon=N_B+\int_{|E|\geq\Delta_0}(\rho_S-\rho_0)d\epsilon.
$$ 
Inserting here Eqs. (\ref{rho}) and (\ref{rho_0}) and using the energy $E$ and the direction of propagation $p=L,R$ to label the scattering states, 
we arrive at the following expression for the number of the bound states:
\begin{equation}
\label{Nb-gen}
  N_B=-\int_{|E|\geq\Delta_0} dE\lim_{\ell\to\infty}\int_{-\ell/2}^{\ell/2}dx\sum_{p=L,R}\left[\tr\,\tilde\psi_{E,p}^\dagger(x)\tilde\psi_{E,p}(x)-\tr\,\psi_{E,p}^{(0),\dagger}(x)\psi_{E,p}^{(0)}(x)\right].
\end{equation}
Since the eigenstates of $H$ and $\tilde H$ are related by a unitary transformation, $\tilde\psi_{E,p}(x)=e^{-i\varphi(x)\hat\sigma_3/2}\psi_{E,p}(x)$, one can replace $\tilde\psi_{E,p}(x)$ in Eq. (\ref{Nb-gen}) by $\psi_{E,p}(x)$. 

The right-hand side of Eq. (\ref{Nb-gen}) is ill-defined because it contains the difference between two infinite ($\delta$-function) normalization integrals for the scattering states with and without the DW. 
To make sense of this expression, we split the energies of the eigenfunctions as follows: $\psi_{E,p}^\dagger\psi_{E,p}-\psi^{(0),\dagger}_{E,p}\psi^{(0)}_{E,p}=[\psi_{E,p}^\dagger\psi_{E',p}-\psi^{(0),\dagger}_{E,p}\psi^{(0)}_{E',p}]_{E'\to E}$.
Then, one can use the identity (\ref{psi12-identity}) to express the integral over coordinates in Eq. (\ref{Nb-gen}) in terms of the asymptotic values of the eigenfunctions far from the DW. In this way, we obtain
\begin{equation}
\label{Nb}
  N_B=-iv_{F,x}\int_{|E|\geq\Delta_0} dE\lim_{\ell\to\infty}P,
\end{equation}
where
$$
  P=\lim_{E'\to E}\frac{1}{E-E'}\sum_{p=L,R}\left[\left.\tr(\psi_{E,p}^\dagger\hat\sigma_3\psi_{E',p})\right|_{-\ell/2}^{\ell/2} 
  -\left.\tr(\psi_{E,p}^{(0),\dagger}\hat\sigma_3\psi_{E',p}^{(0)})\right|_{-\ell/2}^{\ell/2}\right].
$$
The asymptotics of $\psi_{E,p}$ are given by Eqs. (\ref{EL}) and (\ref{ER}), while the expressions for $\psi_{E,p}^{(0}$ can be obtained from Eqs. (\ref{EL}) and (\ref{ER}) by setting $t_{L,R}=1$ and $r_{L,R}=0$. 
After some straightforward algebra, we have
\begin{equation}
\label{P-def}
  P=\lim_{E'\to E}\frac{1}{E-E'}\left[A(E,E')e^{-i(q-q')\ell/2}+B(E,E')e^{i(q+q')\ell/2}+B^*(E',E)e^{-i(q+q')\ell/2}\right],
\end{equation}
where
\begin{eqnarray*}
  && A(E,E')=[t_L^*(E)t_L(E')+r_R^*(E)r_R(E')-1]a_{++}(E,E')-[r_L^*(E)r_L(E')+t_R^*(E)t_R(E')-1]a_{--}(E,E'),\\
  && B(E,E')=r_R(E')a_{-+}(E,E')-r_L(E')a_{+-}(E,E'),
\end{eqnarray*}
and
$$
  a_{ss'}(E,E')=\frac{1}{2}C(E)C(E')\biggl[\sqrt{\Bigl(1+s\frac{v_{F,x}q}{E}\Bigr)\Bigl(1+s'\frac{v_{F,x}q'}{E'}\Bigr)}\sign E\sign E'-
					  \sqrt{\Bigl(1-s\frac{v_{F,x}q}{E}\Bigr)\Bigl(1-s'\frac{v_{F,x}q'}{E'}\Bigr)}\biggr],
$$
with $s,s'=\pm$.

It is easy to see that $A(E,E)=0$, due to Eq. (\ref{tr-props-1}), and $B(E,E)=0$. Therefore,
\begin{equation}
\label{P}
  P=-\left.\frac{\partial A(E,E')}{\partial E'}\right|_{E=E'}-\left.\frac{\partial B(E,E')}{\partial E'}\right|_{E=E'}e^{iql}-\left.\frac{\partial B^*(E',E)}{\partial E'}\right|_{E=E'}e^{-iql}.
\end{equation}
In the first term, one has only to differentiate the coefficients in front of $a_{++}$ and $a_{--}$, because $|t_L|=|t_R|=t$ and $|r_L|=|r_R|=\sqrt{1-t^2}$. In the last two terms, one has only to differentiate $a_{-+}$ and $a_{+-}$.
The result looks as follows: $P=P_1+P_2$, where
\begin{equation}
\label{P1}
  P_1=-\frac{1}{2\pi v_{F,x}}\sign E\left(t_L^*\frac{\partial t_L}{\partial E}+r_L^*\frac{\partial r_L}{\partial E}+t_R^*\frac{\partial t_R}{\partial E}+r_R^*\frac{\partial r_R}{\partial E}\right)
\end{equation}
and
$$
  P_2=-\frac{i}{v_{F,x}}C^2\frac{\Delta_0}{|E|}\im\left[(r_L+r_R)\frac{e^{iql}}{q}\right].
$$
In $P_1$, one can use the definition of the $S$-matrix, Eq. (\ref{S-matrix}), to represent the expression in the brackets as $\tr(S^\dagger\partial S/\partial E)$.
In $P_2$, we observe that in the limit $\ell\to\infty$ only small $q$ are important, corresponding to the energies close to the buk gap edge. For these energies,
the asymptotics (\ref{tr-band edge}) hold and we have $r_L+r_R\to-2$. Therefore,
\begin{equation}
\label{P2}
  \lim_{\ell\to\infty}P_2=\frac{2i}{v_{F,x}}C^2\frac{\Delta_0}{|E|}\lim_{\ell\to\infty}\frac{\sin q\ell}{q}=\frac{i}{v_{F,x}}\left[\delta(E+\Delta_0)+\delta(E-\Delta_0)\right].
\end{equation}
Inserting Eqs. (\ref{P1}) and (\ref{P2}) into Eq. (\ref{Nb}), we obtain 
\begin{equation}
\label{Nb-tr}
  N_B=1+\frac{i}{2\pi}\int_{|E|\geq\Delta_0}dE\sign E\tr\left(S^\dagger\frac{\partial S}{\partial E}\right).
\end{equation}
The unity on the right-hand side originates from the states located exactly at the bulk gap edge, which give rise to the $\delta$-functions in Eq. (\ref{P2}). When integrated over energy, each of these $\delta$-functions contributes 1/2 to $N_B$.
Finally, using $\tr(S^\dagger\partial S/\partial E)=\partial(\ln\det S)/\partial E$, we arrive at Eq. (\ref{Nb-final}).

\end{document}